# Photophysical characterizations of 2-(4-Biphenylyl)-5 phenyl-1,3,4- oxadiazole in restricted geometry

P. K. Paul, S. A. Hussain and D. Bhattacharjee*

Department of Physics, Tripura University,

Suryamaninagar-799130, Tripura, INDIA

**Abstract:**
Langmuir and Langmuir-Blodgett (LB) films of nonamphiphilic 2-(4-Biphenylyl)-5 phenyl-1,3,4- oxadiazole (abbreviated as PBD) mixed with stearic acid (SA) as well as also with the inert polymer matrix poly(methyl methacrylate) (PMMA) have been studied. Surface pressure versus area per molecule ($\pi$-A) isotherms studies suggest that PBD molecules very likely stand vertically on the air-water interface and this arrangement allows the PBD molecules to form stacks and remain sandwiched between SA/PMMA molecules. At lower surface pressure phase separation between PBD and matrix molecules occur resulting due to repulsive interaction. However at higher surface pressure PBD molecules form aggregates. The UV-Vis absorption and Steady state fluorescence spectroscopic studies of the mixed LB films of PBD reveal the nature of the aggregates. H-type aggregates predominates in the mixed LB films whereas I-type aggregates predominates in the PBD-PMMA spin coated films. The degree of deformation produced in the electronic levels are largely affected by the film thickness and the surface pressure of lifting.

* Corresponding author
E.mail: tuphysic@sancharnet.in
       sa_h153@hotmail.com
Phone: +91-381-2375317
Fax: +91-381-2374802

**Introduction:**

Much of the wealth of research in Langmuir-Blodgett films over the last few years has been driven by the technological potential thought to lie in such systems. Ultrathin films such as Langmuir-Blodgett films and Layer-by-Layer self assembled films, have their potential use in the construction of molecular devices [1,2]. Most of the envisioned applications require incorporation of organic molecules having interesting chromophores. Studies on non-linear optical properties [3,4], ferroelectricity [5], pyroelectric properties [6,7], photochromic effect [8] etc have been done on LB films. The well known restricted geometry of the LB film has the advantage of obtaining highly organized supramolecular assemblies at the molecular level [9]. Ideal amphiphilic molecules can form stable Langmuir monolayer at the air-water interface. Non-amphiphilic molecules are not ideally suited for thin film deposition by this technique. However some recent work showed that highly organized LB films of these materials can be formed, when a long chain fatty acid or some inert polymer matrix was used as the supporting medium. The photophysical characteristics of such LB films can be manipulated with ease by changing various LB parameters.

Some of the earlier works [10-14] demonstrated the spectroscopic characterstics of several non-amphiphilic molecules incorporated into the mixed LB films using optically inert matrices as the building block materials.

This communication reports the behaviour of the mixed LB films of non-amphiphilic 2-(4-Biphenylyl)-5 phenyl-1,3,4- oxadiazole (abbreviated as PBD) with a long chain fatty acid namely stearic acid (SA) as well as also with the inert polymer matrix poly(methylmethacrylate) (PMMA) at the air-water interface and their photophysical characteristics when transferred onto the quartz substrates as the LB films.. This low molecular weight electron-deficient (n type) [15] organic material is suitable for incorporation into the organic electroluminescent devices. In this regard aromatic nitrogen containing heterocycles have proved to be particularly effective as electron transporting, hole-blocking (ETHB) [16] materials and in some cases also serving as the emitting component [Ref-10]. These molecules also have high thermal stability and high photoluminescence quantum yield and may be used in scintillator, solar energy concentrator, UV Laser dye [17-18] etc.

**Experimental:**

PBD was purchased from Aldrich chemical company, USA and vacuum sublimed followed by repeated recrystalization before use. SA (purity > 99%) from Sigma, USA and isotactic PMMA from Polyscience were used as received. The solvent chloroform (SRL, India) was of spectral grade and its fluorescence spectrum was checked before use. A commercially available Langmuir Blodgett film deposition instrument (Apex, 2000C, India) was used for mono- and multilayer film depositions. The subphase used was triple distilled deionised water. The pH of the subphase was 6.5 in equilibrium with atmospheric carbondioxide. Solutions of PBD, PMMA, SA as well as PBD-PMMA and PBD-SA mixtures at different molefractions were prepared in chloroform solvent and were spread on the subphase by a micro-syringe. Surface pressure was recorded by using a Wilhelmy plate arrangement as described elsewhere [19]. Deposition of multilayer was achieved by allowing the substrate to dip with a speed of 5 mm/min with a drying time of 15 minute after each lift. Fluorescence grade quartz slides were used for spectroscopic measurements. For each molefraction of PBD, 10 bi-layer LB films were deposited. We chose 15 mN/m as the standard surface pressure for lifting the mixed LB films for both the matrices. The transfer ratio was found to be 0.98 ± 0.02. The Spin Coated films were prepared using a Spin Coating Unit (SCU 2005A, APEX Instrument Co, India) keeping the motor speed fixed at 6000 rpm for a duration of 60 second.

Fluorescence spectra and UV-Vis absorption spectra were measured by a Perkin Elmer LS 55 spectrophotometer and Perkin Elmer Lambda 25 spectrophotometer respectively. All the measurements were performed at room temperature ($24^O$C).

**Pressure-area isotherm:**

Isotherm characteristics of pure PBD were studied by spreading dilute solution of pure PBD in Chloroform ($2\times10^{-3}$M) on the water surface of the Langmuir trough. After sufficient time was allowed to evaporate the solvent, the barrier was compressed very slowly at a speed of $2\times10^{-3}$ $nm^2$ $mole^{-1}s^{-1}$. It was observed that the surface pressure rose up to 27mN/m. However, the area per molecule of pure PBD was found to be negligibly small in comparison to the actual planar area of the molecules as predicted by the space filling model. This indicates that the pure PBD molecules do not form monolayer at the air-water interface rather multilayer or aggregates of pure PBD is formed. The transfer of this floating layer onto a solid substrate produced a patchy and inhomogeneous film. However when PBD was mixed with a supporting matrix of either SA or PMMA a highly stable compressible monolayer at the air water interface was obtained, as was evidenced from the isotherm characteristics. This monolayer could be easily transferred onto solid substrates with a good transfer ratio. The mono and multilayered LB films thus formed were homogeneous and uniform.

Figures 1a and 1b show the surface pressure $(\pi)$ versus area per molecule $(A)$ isotherms of PBD mixed with PMMA and SA respectively at different molefractions of PBD along with pure PMMA and SA isotherms. The area per molecules of pure PMMA and SA are 0.11 and 0.23 $nm^2$, respectively at a surface pressure of 15 mN/m and that at 25 mN/m is 0.21$nm^2$ for pure SA which is consistent with the reported results [12].

Pure PMMA isotherm shows an inflection point at about 20 mN/m [20]. it is also observed that this inflection point gradually loses its distinction with increasing molefractions of PBD in PMMA and at and above 0.7M of PBD, the PBD-PMMA mixed isotherms show steep rising without any transition point. Also above this inflection point, the collapsing of monolayer occurs and the monolayer remains no longer stable [20]. This is why a surface pressure of 15mN/m has been chosen for film deposition.
Also from the PBD-PMMA mixed isotherms (figure 1a) it is observed that at lower surface pressures area per molecule decreases consistently with increasing mole fraction. The consistent decrease in the area per molecule with the increase in mole fractions of PBD indicates repulsive interaction between PBD and PMMA molecules that may originate from the completely different physical and chemical characteristics of the PBD and PMMA molecules.

The isotherm characteristics of PBD-SA (shown in figure 1b) mixed monolayer also show same characteristics and is an indication of phase separation between PBD and SA molecules resulting due to repulsive interaction.

Furthermore, the low area per molecule obtained at the air-water interface compared to the large area per molecule for PBD as predicted by the space filling model, suggests that the PBD moieties do not lay flat at the air-water interface but very likely stand vertically on their edges. Moreover, this conclusion receives further support from the fact that such an arrangement allows the possibility for the molecules to form stacks and remain sandwiched in between SA/PMMA molecules. Spectroscopic studies of the LB film as discussed in a latter section of this paper confirm such an arrangement of the molecules in the LB films. Additionally, the plot of absorbance and emission versus concentration of PBD and the number of layers (figure not shown) is found to be almost linear, which confirms that PBD molecules not only are incorporated into the LB layers but also are reproducibly transferred onto the substrates.

Spectroscopic characterization:

The UV-Vis absorption and steady state fluorescence spectra of mixed LB films of PBD in PMMA and SA matrices, respectively, at different molefractions of PBD in PMMA/SA matrices along with the PBD solution and microcrystal spectra for comparison are shown in figures 2a and 2b respectively. Pure PBD solution absorption spectrum shows distinct band systems in the 225 – 350 nm region having distinct 0 – 0 band at 305 nm and a weak hump, which extends from 240 – 265 nm. The microcrystal absorption spectrum shows an overall broadening of the band system having a broad band with peak at around 288 nm along with almost indistinguishable weak hump at around 240-265 nm region. The LB film absorption spectra for various molefractions of PBD in both the matrices are almost similar to microcrystal spectrum having almost identical band positions and broadened spectral profile with respect to pure PBD solution absorption spectrum. It is interesting to note that for all the molefractions of PBD in both the matrices the 0-0 band at 288 nm is blue shifted and broadened with respect to the solution absorption spectrum. Also for lower molefraction of PBD in SA matrix (namely 0.1-0.3M) the vibrational band systems are too weak to be distinguished. The broadening of the 0-0 band along with the blue shift may be due to aggregation.

It is relevant to mention in this context that according to the intermediate strength exciton coupling theory [21,22], dipole-dipole interaction results in the raising or lowering of the exciton band to a position either energetically higher or lower than the monomer band. Such a change in energy is given by;

$$\Delta E = \frac{2M^2(1-3\cos^2\theta)(1-1/N)}{r^3}$$

where M is the dipole moment vector, N is the number of monomers in the aggregate, $\theta$ is the angle between the dipole moment of the molecule and r, the vector joining the centers of two dipoles. When $0^0 < \theta < 54.7^0$, the exciton band is energetically located below the monomeric band that causes a red shift and the corresponding aggregates are referred to as the J-aggregates, while for $54.7^0 < \theta < 90^0$, the exciton band is located energetically above the monomeric band that causes a blue shift and the corresponding aggregates are referred to as H-aggregates. Corresponding to the magic angle $\theta = 54.7^0$, no shift in the absorption spectrum is observed and corresponding aggregates are referred to as I-aggregates.

Here for both the matrices the broadening of the absorption bands accompanied with the blue shift in the mixed LB films of PBD seems to be due to the formation of H-type of aggregates.

It may be mentioned in this context that there are certain rigid nearly planar plate like molecules namely, pyrene [23] or rod like molecules anthracene [24], biquinoline [10] etc., form aggregate in the mixed LB films. These molecules are sandwiched among the matrix molecules (SA/PMMA) to form aggregates in the LB films and partial or total binary demixing is occurred in the mixed LB films. In the present work, PBD molecule has almost linear rod like structure. For aggregation to be occurred in the mixed LB films of PBD molecules, the most possibility is the sandwich of PBD molecules among SA or PMMA molecules in the mixed LB films.

Pure PBD solution fluorescence spectrum in chloroform (figure 2a and 2b) shows a distinct and prominent band profile within 325-400 nm region with prominent high energy O-O band at 342 nm along with other prominent intense bands at 358 and 380 nm. There is a large shift of the O-O band at about 38 nm in pure PBD solution fluorescence spectrum in comparison with the absorption spectrum. This may be due to the deformation produced in the electronic states of the molecules. The fluorescence spectrum of PBD microcrystal is somewhat different from the solution spectrum. The high energy bands at 342 nm and 358 nm in solution fluorescence spectrum are totally absent in the microcrystal spectrum having its O-O band at about 374 nm along with prominent band at 391 nm. Also the spectral profile is overall broadened.

Almost all the bands of PBD microcrystal spectrum are present in the fluorescence spectra of the mixed LB films of PBD in PMMA and SA matrices at different mole fractions of PBD (figure 2a and 2b) having O-O band at 374 nm. Along with the band at 391 nm, there is a weak hump at around 418 nm in the longer wavelength region of the fluorescence spectra of the mixed LB film, which was absent in the PBD microcrystal spectrum. The quenching of the high energy bands may be due to the reabsorption effect arising due to the formation of aggregates and have been observed in several other molecules [25].

Figure 3 shows the UV-Vis absorption and steady state fluorescence spectra of PBD-PMMA spin coated films at different mole fractions of PBD in PMMA. Here the absorption spectra show diffused, broadened spectral profile and up to 0.7 mole fraction of PBD there is almost no shifting of band position with respect to the pure PBD solution spectrum. The broadening of the absorption band accompanied with no shift in the PBD-PMMA mixed spin coated films may be due to the formation of I-type of aggregates, where long axis of the neighbouring PBD molecules are parallel to each other and makes an angle of ~ 57° with the vector joining the centers of the PBD molecules.

It is interesting to mention in this context that for J or H type of aggregate, the molecules are closely packed and have definite relative orientations and dictated by the surfaces and the staking. In case of spin coated film, the molecules are in a closely packed and I-type aggregates are occurred. So broadening only manifests the absorption spectra of PBD-PMMA mixed spin coated films with almost no shifting. But the LB technique provides a highly ordered environment for the molecules where specific type of aggregate (H-type) of PBD molecule can occur in a controlled manner.

Figure 3 also shows the fluorescence spectra of PBD-PMMA mixed spin coated films at different molefractions of PBD in PMMA. Unlike that of mixed LB films of PBD fluorescence spectra, here for lower molefractions the band profile is overall broadened and diffused and does not contain any prominent vibrational pattern. However for higher molefraction the fluorescence spectra have distinct similarity with the microcrystal and mixed LB film fluorescence spectra.

**Layer Effect:**

Various technical applications sometimes demand thick films. Here we have studied the dependence of photophysical characteristics of mixed LB films of PBD with thickness in the light of UV-Vis absorption and Steady state fluorescence spectroscopy. Figures 4a, 4b and 5a, 5b show the UV-Vis absorption and Steady state fluorescence spectra of different layered mixed LB films of PBD in PMMA and SA matrices respectively at two different molefractions of 0.1M and 0.5M of PBD.

UV-Vis absorption spectra of different layered mixed LB films of PBD in PMMA and SA for both the molefractions show almost identical spectral profile except a small change in intensity distribution. However, their fluorescence spectra show quite interesting features. The fluorescence spectra of 0.1 M of PBD in PMMA (figure 4a) show that there exist a weak hump at high energy region with peak at 359 nm alongwith a broad longer wavelength band at 440 nm. These two bands are absent at higher molefraction of PBD in PMMA as well as also in SA matrix. This high energy band has distinct similarity with the intense prominent band at 358 nm in PBD solution spectrum.

It may be worthwhile to mention in this context that PMMA molecules have coil like [20] structure. At lower molefraction of PBD in PMMA, the amount of PBD molecules are small in the PBD-PMMA mixed films and the PBD molecules find an excellent microenvironment where they can accommodate easily and get isolated from each other. As a result the deformation produced in the electronic levels may be reduced slightly resulting the appearance of high energy band in the fluorescence spectra. With the increase in layer number, this band becomes prominent and distinct. However at higher molefraction of PBD in PMMA, the amount of PBD molecules being large in the PBD-PMMA mixed films, the packing pattern and aggregation of PBD molecules are different resulting in the disappearance of high energy band.

**Pressure Effect:**

The morphology and crystal parameter as well as the packing pattern of molecules can be controlled by controlling the surface pressure of lifting of LB films, which may suit for any specific technical applications.

Here we have studied the dependence of photophysical characteristics of PBD-SA mixed LB films lifted at different surface pressures. Pressure effect studies of PBD-PMMA mixed LB films are not possible as PMMA monolayer as well as PBD-PMMA monolayer at the air-water interface remain no longer stable beyond 18 mN/m surface pressure [20]. Figure 6a and 6b show the UV-Vis absorption and steady state fluorescence spectra of PBD-SA mixed LB films at two different molefractions 0.1 and 0.5M of PBD in SA respectively, and lifted at 15, 20, 25, 30 and 35 mN/m surface pressure along with the solution and microcrystal spectra for comparison.

The absorption spectra of mixed LB films of 0.1 M of PBD in SA matrix lifted at different surface pressures (figure 6a) are extremely weak with almost indistinguishable bands. However, the absorption spectra of 0.5 M PBD-SA mixed LB films show prominent bands (figure 6b) and have distinct similarity with that of microcrystal spectrum.

The fluorescence spectra of 0.1 M PBD-SA mixed LB films give interesting features at higher surface pressures of 30 and 35 mN/m. At these surface pressures prominent high energy band at around 354 nm is observed and this band has distinct similarity with the high energy solution fluorescence band at 358 nm. This band is absent in the microcrystal spectrum. This high energy band at 354 nm is totally absent in the fluorescence spectra of 0.5 M of PBD-SA mixed LB films lifted at higher surface pressures as well as in all other surface pressures.

As mentioned earlier the PBD molecules are stacked among the SA molecules. With the increase in surface pressure, the packing patterns of PBD molecules in the PBD-SA mixed LB films are changed. Especially, at lower molefractions of PBD, due to non-availability of sufficient amount of PBD molecules, the stacking pattern may be different and at higher surface pressure the deformation produced in the electronic level is partially lifted, as a result, high energy band at 354 nm occurs. However with the increase in molefraction, closer association of PBD molecules occurs resulting in the absence of the high energy band.

**Conclusion:**

In conclusion our observations show that non-amphiphilic 2-(4-Biphenylyl)-5 phenyl-1,3,4- oxadiazole (abbreviated as PBD) form well organized stable Langmuir films at the air-water interface when mixed with a long chain fatty acid namely stearic acid (SA) as well as also with the inert polymer matrix poly(methyl methacrylate) (PMMA). Stable mono- and multilayer Langmuir-Blodgett (LB) films can be obtained by transferring this Langmuir monolayer onto quartz substrates. Surface pressure versus area per molecule ($\pi$-A) isotherm studies reveal that the PBD moieties do not lie flat at the air-water interface but very likely stand vertically on their edges. At lower surface pressure phase separation between PBD and matrix molecules occur resulting due to the repulsive interaction. However at higher surface pressure PBD molecules form aggregates. The UV-Vis absorption and Steady state fluorescence spectroscopic studies of the mixed LB films of PBD confirm the formation of H-type of aggregates in the mixed LB films whereas I-type aggregates predominates in the PBD-PMMA spin coated films. The degree of deformation produced in the electronic levels are largely affected by the film thickness, molefraction of mixing and the surface pressure of lifting.

**Acknowledgement:**

The authors are grateful to DST and CSIR, Govt. of India for providing financial assistance through FIST-DST Project No. SR/FST/PSI-038/2002 and CSIR project Ref. No. 03(1080)/06/EMR-II

**Figure Captions**

Fig: 1(a): Surface pressure ($\pi$) versus area per molecule (A) isotherms of PBD in PMMA matrix at different molefractions of PBD. The numbers denote corresponding molefractions of PBD in PMMA matrix. PMMA and PBD correspond to the pure PMMA and pure PBD isotherms. The inset shows the molecular structure of PBD.

Fig: 1(b): Surface pressure ($\pi$) versus area per molecule (A) isotherms of PBD in SA matrix at different molefractions of PBD. The numbers denote corresponding molefractions of PBD in SA matrix. SA and PBD correspond to the pure SA and pure PBD isotherms

Fig: 2(a): UV-Vis absorption and steady state fluorescence spectra of PBD in chloroform solution (CHCl$_3$), in microcrystal (MC) and in PBD/PMMA mixed LB films. The numbers denote the corresponding molefractions of PBD in PMMA matrix.

Fig: 2(b): UV-Vis absorption and steady state fluorescence spectra of PBD in chloroform solution (CHCl$_3$), in microcrystal (MC) and in PBD/SA mixed LB films. The numbers denote the corresponding molefractions of PBD in SA matrix.

Fig: 3: UV-Vis absorption and steady state fluorescence spectra of PBD in chloroform solution (CHCl$_3$), in microcrystal (MC) and PBD/PMMA spin coated films. The numbers denote the corresponding molefractions of PBD in SA matrix.

Fig: 4(a): UV-Vis absorption and steady state fluorescence spectra of different layer mixed LB films of PBD in PMMA matrix at 0.1 M of PBD in PMMA. The numbers denote the corresponding number of layer.

Fig: 4(b): UV-Vis absorption and steady state fluorescence spectra of different layer mixed LB films of PBD in PMMA matrix at 0.5 M of PBD in PMMA. The numbers denote the corresponding number of layer.

Fig: 5(a): UV-Vis absorption and steady state fluorescence spectra of different layer mixed LB films of PBD in SA matrix at 0.1 M of PBD in SA. The numbers denote the corresponding number of layer.

Fig: 5(b): UV-Vis absorption and steady state fluorescence spectra of different layer mixed LB films of PBD in SA matrix at 0.5 M of PBD in SA. The numbers denote the corresponding number of layer.

Fig: 6(a): UV-Vis absorption and steady state fluorescence spectra of PBD/SA mixed LB films at different surface pressures of lifting at a molefractions of 0.1 M of PBD in SA. The numbers denote the corresponding surface pressure of lifting and molefraction.

Fig: 6(b): UV-Vis absorption and steady state fluorescence spectra of PBD/SA mixed LB films at different surface pressures of lifting at a molefractions of 0.5 M of PBD in SA. The numbers denote the corresponding surface pressure of lifting and molefraction.

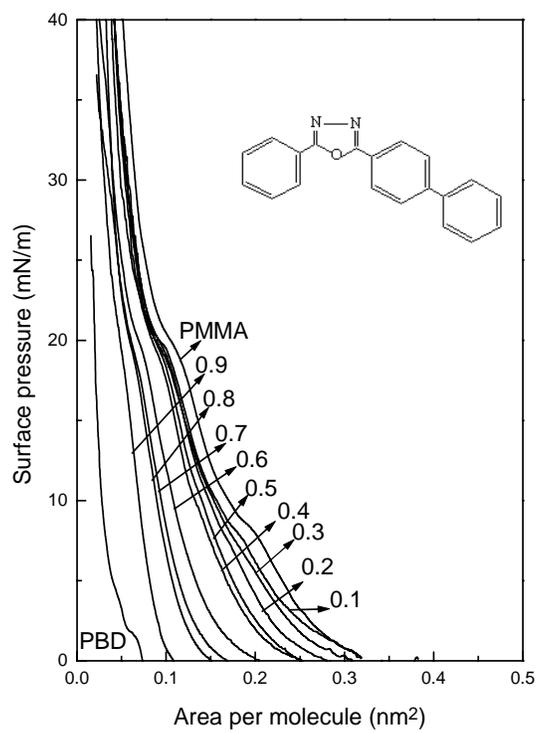

Figure 1(a) P. K. Paul et. al.

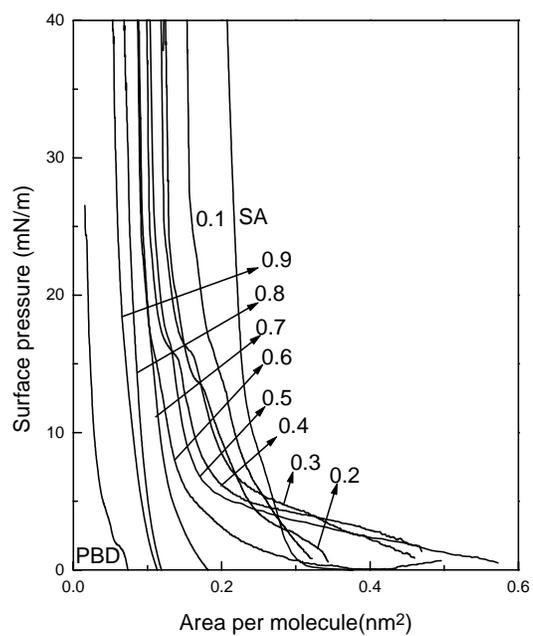

Figure 1(b) P. K. Paul et. al.

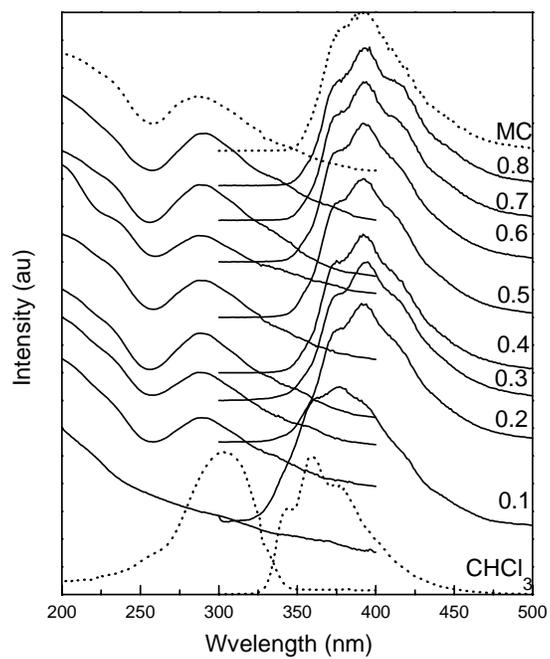

Figure 2(a) P. K. Paul et. al.

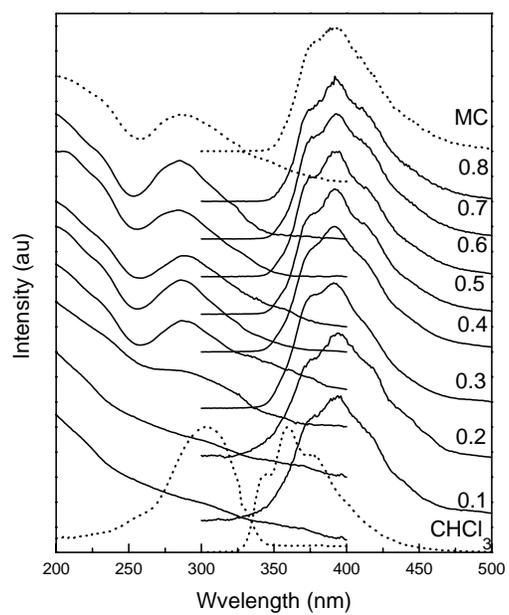

Figure 2(b) P. K. Paul et. al.

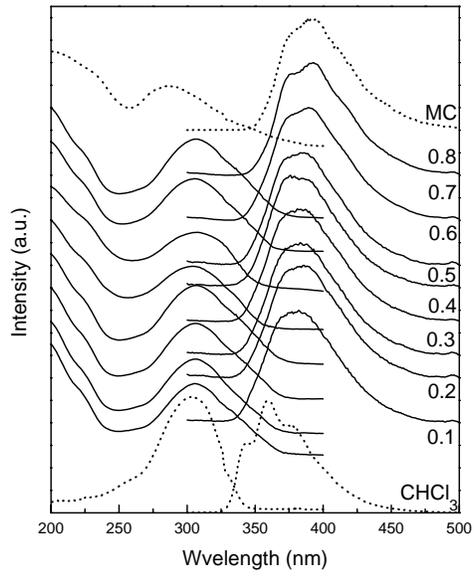

Figure 3 P. K. Paul et. al.

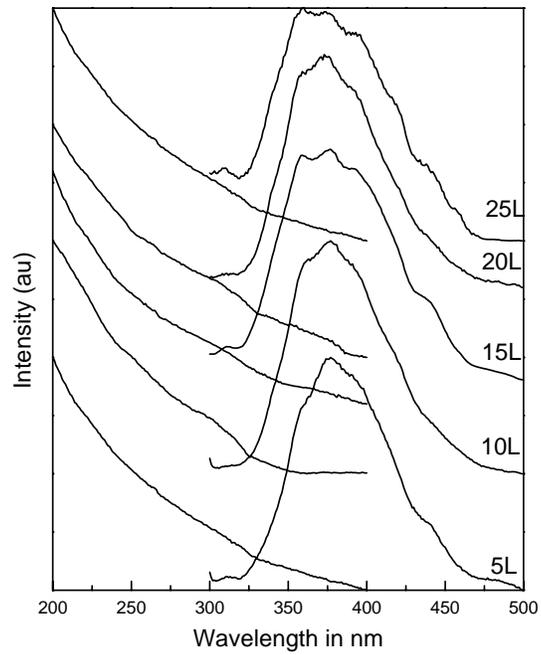

Figure 4(a) P. K. Paul et. al.

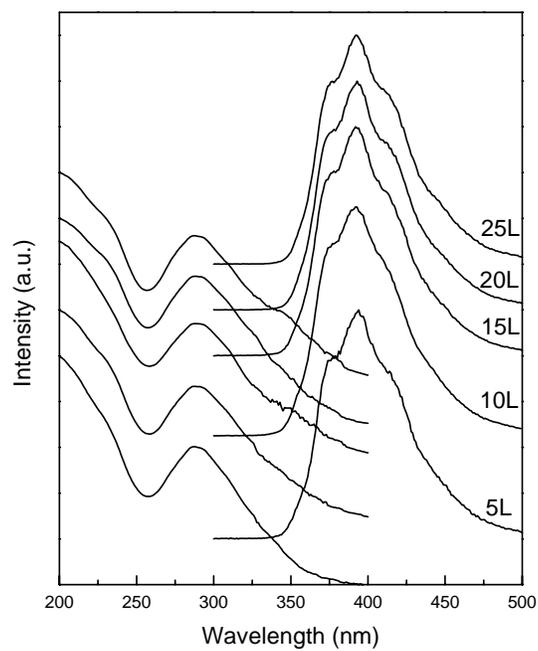

Figure 4(b) P. K. Paul et. al.

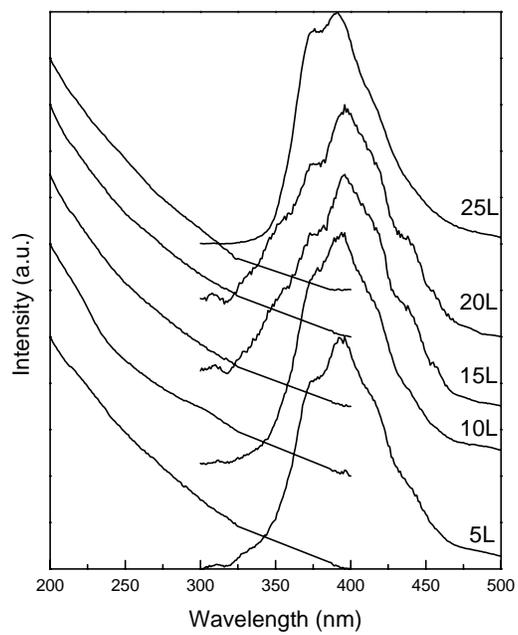

Figure 5(a): P. K. Paul et. al.

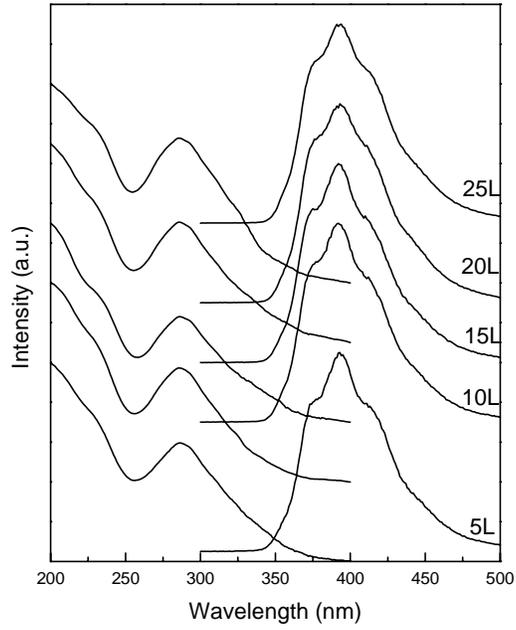

Figure 5(b) P. K. Paul et. al.

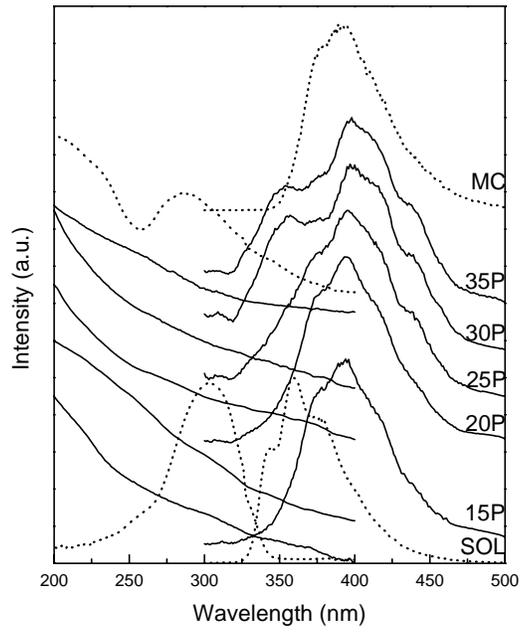

Figure 6(a) P. K. Paul et. al.

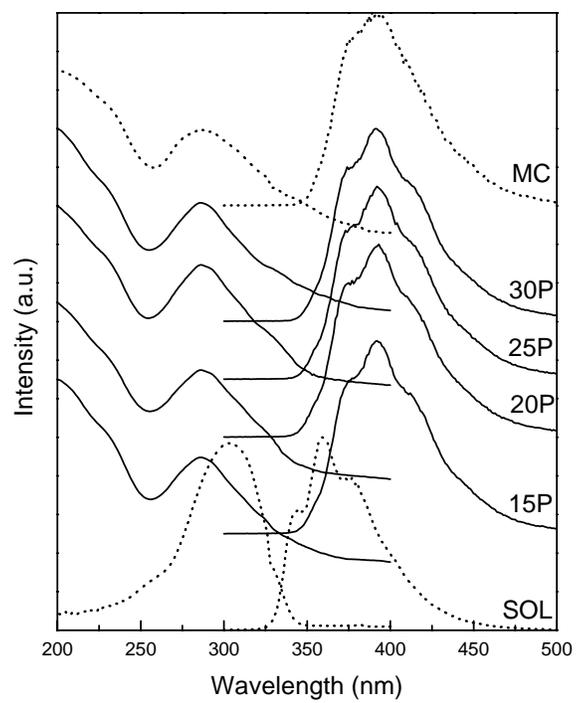

Figure 6(b) P. K. Paul et. al.